\documentclass[useAMS,usegraphicx,usenatbib]{mn2e}
\usepackage{times}

\newcommand{\Msun}{\ensuremath{\,{\rm M}_\odot}}                        
\newcommand{\Rsun}{\ensuremath{\,{\rm R}_\odot}}                        
\newcommand{\Lsun}{\ensuremath{\,{\rm L}_\odot}}                        
\newcommand{\Teff}{\ensuremath{T_{\rm eff}}}                            
\newcommand{\logg}{\ensuremath{\log g}}                                 
\newcommand{\Mbol}{\ensuremath{M_{\rm bol}}}                            
\newcommand{\EBV}{\ensuremath{E_{B-V}}}                                 
\newcommand{\kms}{\,km\,s$^{-1}$}                                       
\newcommand{\Vsini}{\ensuremath{V_{\sin\!i}}}                           
\newcommand{\Vsync}{\ensuremath{V_{\rm synch}}}                         
\newcommand{\mc}[1]{\multicolumn{2}{c}{#1}}
\newcommand{\reff}[1]{{#1}}                                         

\title[The eclipsing binary system XY\,Ceti]
      {Absolute dimensions of detached eclipsing binaries. II. The metallic-lined system XY\,Ceti}

\author[Southworth et al.]
       {John Southworth$^1$\thanks{E-mail: jkt@astro.keele.ac.uk},
        K.\ Pavlovski$^{1,2}$, E.\ Tamajo$^2$, B.\ Smalley$^1$, R.\ G.\ West$^3$, D.\ R.\ Anderson$^1$ \\
        $^1$\,Astrophysics Group, Keele University, Staffordshire, ST5 5BG, UK \\
        $^2$\,Department of Physics, University of Zagreb, Bijeni\v{c}ka cesta 32, 10000 Zagreb, Croatia \\
        $^3$\,Department of Physics and Astronomy, University of Leicester, University Road, Leicester LE1 7RH
        }

\begin{document} \maketitle 

\begin{abstract}
We present phase-resolved spectroscopy and extensive survey photometry of the detached eclipsing binary system XY\,Cet, which is composed of two metallic-lined stars. We measure their masses to be $1.773 \pm 0.016$ and $1.615 \pm 0.014$\Msun\ and their radii to be $1.873 \pm 0.035$ and $1.773 \pm 0.029$\Rsun, resulting in logarithmic surface gravities of $4.142 \pm 0.016$ and $4.149 \pm 0.014$ (cgs). We determine effective temperatures of $7870 \pm 115$ and $7620 \pm 125$ K. The projected rotational velocities are $34.4 \pm 0.4$ and $34.1 \pm 0.4$\kms, which are close to synchronous. Theoretical models cannot match all of these properties, but come closest for a solar helium and metal abundance and an age in the region of 850\,Myr. We obtain the individual spectra of the two stars by the spectral disentangling method, and compare them to synthetic spectra calculated for the measured effective temperatures and a solar chemical composition. Both stars show enhanced abundances of iron-group elements and clear deficiencies of Ca\,I and Sc\,II, confirming their classification as Am stars. We also find strong overabundances of Zr\,II and the rare-earth species La\,II, Ce\,II and Nd\,II, a hallmark of chemically peculiar A\,stars. XY\,Cet is a prime candidate for detailed spectroscopic analyses of metallic-lined stars whose masses and radii are known to accuracies of 1--2\%.
\end{abstract}



\section{Introduction}                                                                                                               \label{sec:intro}

\reff{Eclipsing binary star systems are of immense value as our primary source of empirical measurements of the properties of stars \citep{Andersen91aarv,Torres++10aarv}. The masses and radii of those systems which show double-lined spectra can be measured using purely geometrical arguments and to precisions of better than 1\% \citep[e.g.][]{Clausen+09aa}. With effective temperature determinations they become excellent distance indicators both within the Milky Way and to external galaxies \citep{Ribas+05apj,Bonanos+06apj}. Detached eclipsing binaries (dEBs) are the pick of the bunch as their properties can be used to challenge stellar theory and single-star evolutionary models.}

One use of dEBs is to probe chemical peculiarities, such as the Am phenomenon in stars \citep{TitusMorgan40apj,Conti70pasp} by measuring their physical properties to high accuracy. Am stars are A\,stars which show photospheric abundance anomalies thought to be caused by radiative diffusion and gravitational settling \citep{,Michaud70apj,Turcotte+00aa,Talon++06apj}. These effects are able to operate in the radiative atmospheres possessed by A\,stars with rotational velocities slower than roughly 100\kms\ \citep{AbtLevy85apjs,Budaj96aa,Budaj97aa}. Am stars are preferentially found in short-period binaries where tidal effects have been able to slow their rotation \citep{Abt61apjs,Abt65apjs,CarquillatPrieur07mn}.

This in turn means that Am stars are strongly represented among the well-studied dEBs\footnote{A catalogue of well-studied detached eclipsing binaries is maintained at {\tt http://www.astro.keele.ac.uk/$\sim$jkt/debdata/debs.html}}, such as $\beta$\,Aur \citep{Me++07aa}, V364\,Lac \citep{Torres+99aj}, V624\,Her \citep{Popper84aj2}, V459\,Cas \citep{Lacy++04aj}, WW\,Cam \citep{Lacy+02aj} and RR\,Lyn \citep{TomkinFekel06aj}. The prevailing viewpoint from these studies is that the Am phenomenon is a surface disease -- the properties of Am stars can in general be matched by theoretical predictions just as well as those of normal A\,stars -- although a high bulk metal abundance (5\% by mass) was found for the component stars of WW\,Aurigae by \citet{Me+05mn}.

XY\,Ceti is a dEB containing Am stars which has been studied on many occasions but whose physical properties were not firmly established. In this work we present new spectroscopy and extensive photometric measurements from the SuperWASP survey from which we measure the physical properties of the component stars to high precision. This study highlights the many possibilities opening up for the study of eclipsing binaries of all types from the increasing number of long-term variability surveys, in particular those which have the photometric precision to search for extrasolar planetary transits.

\subsection{XY Ceti}

\begin{table} \begin{center}
\caption{\label{tab:basedata} Identifications, location, and photometric
indices for XY\,Cet. \newline {\bf References:} (1) \citet{Perryman+97aa};
(2) \citet{Hog+97aa}; (3) \citet{CannonPickering18anhar2};
(4) \citet{Argelander03book}; (5) This work; (6) \citet{Skrutskie+06aj}.}
\begin{tabular}{lr@{}lr} \hline
\                          &   & XY Ceti           & Reference \\
\hline
Hipparcos number           &   & HIP 13937         &       1   \\
Tycho designation          &   & TYC 51-832-1      &       2   \\
Henry Draper number        &   & HD 18597          &       3   \\
Bonner Durchmusterung      &   & BD +02\degr 460   &       4   \\[2pt]
RA (J2000)                 &   & 02 59 33.5333     &       1   \\
Dec.\ (J2000)              & + & 03 31 03.274      &       1   \\
Hipparcos parallax ($m$as) &   & $1.72 \pm 1.52$   &       1   \\
Spectral type              &   & A8 Vm + F0 Vm     &       5   \\[2pt]
$B_T$                      &   & 9.094 $\pm$ 0.024 &       2   \\
$V_T$                      &   & 8.800 $\pm$ 0.026 &       2   \\
$J_{\rm 2MASS}$            &   & 8.212 $\pm$ 0.023 &       6   \\
$H_{\rm 2MASS}$            &   & 8.194 $\pm$ 0.031 &       6   \\
$K_{\rm 2MASS}$            &   & 8.115 $\pm$ 0.021 &       6   \\
\hline \end{tabular} \end{center} \end{table}

The object BD\,+02\degr 460 was discovered to be an eclipsing binary system by \citet{StrohmeierKnigge61}, who found that an orbital period of 1.390356\,d satisfied their photographic observations but yielded a light curve with no detectable secondary minimum. The first photoelectric study of XY\,Cet \citep{MorrisonMorrison68aj} revised the orbital period to 2.780712\,d, resulting in a light curve containing rather similar primary and secondary minima. Morrison \& Morrison analysed their light curve using the method of \citet{RussellMerrill52book}, finding that the two stars were very similar.

\citet{Popper69baas} announced that both components of XY\,Cet had a metallic-lined nature. This was independently discovered by Dr.\ A.\ Young \citep{Popper71apj}. The first -- and so far only -- time-resolved spectroscopic study was presented by \citet{Popper71apj} based on photographic material. Popper combined his spectroscopic results with the photometric elements of \citet{MorrisonMorrison68aj} to make the first determination of the masses and radii of the component stars of XY\,Cet, finding the two objects to have similar properties (masses 1.75 and 1.63\Msun\ and radii 1.95 and 1.89\Rsun). Popper gave a rough spectral classification of A2m + F0m.

However, an analysis of new photometric data for XY\,Cet led \citet{SrivastavaPadalia75apss} to radii which were quite different (2.14 and 1.61\Rsun). \citet{Okazaki89apss} revisited the photometric data obtained by both \citet{MorrisonMorrison68aj} and \citet{SrivastavaPadalia75apss} and resolved this discrepancy in favour of Popper's results (radii 1.96 and 1.78\Rsun), in the process finding that XY\,Cet has a small but significant orbital eccentricity. However, a substantial disparity between the radii of the two components was resurrected by \citet{Ramella+80apss}, through an analysis of the data obtained by \citet{SrivastavaPadalia75apss}. These authors used, for the first time in the case of XY\,Cet, `modern light curve modelling techniques' to find radii of 2.10 and 1.74\Rsun. These values were supported by \citet{Srivastava75apss}, who reanalysed the data from \citet{SrivastavaPadalia75apss} to find values of 2.13 and 1.76\Rsun.

Since the apparent resolution of varied measurements of the radii of the component stars of XY\,Cet in favour of a notable difference between the two objects, there has been only a little interest in this binary system. \citet{Srivastava87apss} presented a period study of XY\,Cet, finding five changes in period but providing no clear evidence for their existence. Finally, one spectrum of the object was obtained by \citet{Glazunova+08aj}, who measured the rotational velocities of both components. Their observation was taken at an orbital phase of 0.957, where the spectral lines of the two components would be partially overlapping, so it is not surprising that their results are inconsistent with the values we find in Sect.\,\ref{sec:atm}. The basic observational properties of XY\,Cet are given in Table\,\ref{tab:basedata}. Throughout this work we refer to the primary star as star\,A and the secondary star as star\,B. Star\,A is hotter, larger and more massive than star\,B.

\section{Data acquisition}                                                                                                             \label{sec:obs}

\subsection{Spectroscopy}

Spectroscopic observations were obtained in 2002 October using the 2.5\,m Isaac Newton Telescope (INT) on La Palma. The 500\,mm camera of the Intermediate Dispersion Spectrograph (IDS) was equipped with a holographic 2400\,lines\,mm$^{-1}$ grating. An EEV 4k\,$\times$\,2k CCD was used and exposure times were 300\,s. From measurements of the full width at half maximum (FWHM) of arc lines taken for wavelength calibration we estimate that the resolution is 0.2\,\AA\ (1.8 pixels). A total of 38 spectra were taken covering the interval 4230--4500\,\AA, with estimated signal to noise ratios of roughly 150 per pixel. One spectrum was taken covering 4710--4970\,\AA, aimed at the H$\beta$ line.

The reduction of all spectra was undertaken using optimal extraction \citep{Horne86pasp} as implemented in the software tools {\sc pamela} and {\sc molly} \footnote{{\sc pamela} and {\sc molly} were written by Dr.\ Tom Marsh and are available at {\tt http://www.warwick.ac.uk/go/trmarsh}} \citet{Marsh89pasp}.

\subsection{Photometry}

Our main photometric dataset comes from the WASP survey for transiting extrasolar planets \citep{Pollacco+06pasp} which operates two instruments, one in La Palma and one in South Africa. Each instrument consists of eight small telescopes on a common mount, and each telescope comprises a Canon 200\,mm telephoto lens, a custom filter\footnote{See {\tt www.superwasp.org}} resembling $g$$+$$r$ \citep{Fukugita+96aj}, and an E2V 2k$\times$2k CCD. Both instruments studied the area surrounding XY\,Cet extensively in the 2008 and 2009 observing seasons. The data were reduced by a dedicated pipeline \citep{Pollacco+06pasp} which performs aperture photometry in three software apertures and then applies the {\sc SysRem} detrending algorithm \citep{Tamuz++05mn} to the photometry from the second of these apertures.

\reff{The WASP observations of XY\,Cet consist of 15\,227 datapoints in total, of which 13\,702 were observed in good conditions during the 2008 and 2009 seasons.} From preliminary light curve fits (see Sect.\,\ref{sec:lc}) we established that the four different photometric aperture/detrending approaches gave results in very good agreement with each other and that the {\sc SysRem}-detrended data had the lowest scatter. We therefore adopted the detrended data for all further analyses. \reff{These data were iteratively clipped to remove outliers at the 4$\sigma$ level, resulting in a final tally of 9111 good datapoints from the 2008 season and 4091 from the 2009 season.}

\begin{table} \begin{center}
\caption{\label{tab:minima} Literature times of minimum light of XY\,Cet and the observed
minus calculated ($O-C$) values of the data compared to the ephemeris derived in this work.
\newline {\bf References:} (1) \citet{StrohmeierKnigge61}; (2) \citet{MorrisonMorrison68aj};
(3) \citet{Baldwin73ibvs}; (4) \citet{Srivastava75apss}; (5) \citet{Kurpinska96conf};
(6) \citet{Locher72bbsag}; (7) \citet{Locher72bbsag2}; (8) \citet{Diethelm74bbsag};
(9) \citet{Diethelm74bbsag2}; (10) \citet{Alnaimiy+78ibvs}; (11) \citet{Pasche89bbsag};
(12) \citet{Pasche89bbsag2}; (13) \citet{Diethelm91bbsag}; (14) \citet{Blattler92bbsag};
(15) \citet{Zejda04ibvs}; (16) This work.}
\begin{tabular}{lrrr} \hline
Time of minimum           &    Cycle    &\hspace*{25pt}$O-C$&  Reference  \\
(HJD $-$ 2\,400\,000)     &    number   &      (HJD)        &             \\
\hline
26734.271  $\pm$ 0.04     &  $-$4185.5  &         0.00724   &   1         \\
26945.596  $\pm$ 0.04     &  $-$4109.5  &      $-$0.00206   &   1         \\
26945.611  $\pm$ 0.04     &  $-$4109.5  &         0.01294   &   1         \\
27365.467  $\pm$ 0.04     &  $-$3958.5  &      $-$0.01895   &   1         \\
27386.362  $\pm$ 0.04     &  $-$3951.0  &         0.02069   &   1         \\
27397.423  $\pm$ 0.04     &  $-$3947.0  &      $-$0.04117   &   1         \\
28127.413  $\pm$ 0.04     &  $-$3684.5  &         0.01127   &   1         \\
36843.514  $\pm$ 0.04     &   $-$550.0  &      $-$0.03740   &   1         \\
36943.559  $\pm$ 0.04     &   $-$514.0  &      $-$0.09812   &   1         \\
36943.604  $\pm$ 0.04     &   $-$514.0  &      $-$0.05312   &   1         \\
36850.510  $\pm$ 0.04     &   $-$547.5  &         0.00681   &   1         \\
36850.556  $\pm$ 0.04     &   $-$547.5  &         0.05281   &   1         \\
36903.379  $\pm$ 0.04     &   $-$528.5  &         0.04224   &   1         \\
37316.292  $\pm$ 0.04     &   $-$380.0  &         0.01913   &   1         \\
38372.949  $\pm$ 0.005    &        0.0  &         0.00461   &   2         \\
39878.701  $\pm$ 0.01     &      541.5  &      $-$0.00030   &   3         \\
40529.387  $\pm$ 0.005    &      775.5  &      $-$0.00150   &   4         \\
40532.161  $\pm$ 0.005    &      776.5  &      $-$0.00821   &   4         \\
40543.298  $\pm$ 0.005    &      780.5  &         0.00593   &   4         \\
40557.194  $\pm$ 0.005    &      785.5  &      $-$0.00164   &   4         \\
40889.495  $\pm$ 0.01     &      905.0  &         0.00397   &   5         \\
40906.177  $\pm$ 0.005    &      911.0  &         0.00168   &   4         \\
40917.301  $\pm$ 0.005    &      915.0  &         0.00283   &   4         \\
40931.204  $\pm$ 0.005    &      920.0  &         0.00225   &   4         \\
40967.357  $\pm$ 0.01     &      933.0  &         0.00597   &   5         \\
40981.270  $\pm$ 0.01     &      938.0  &         0.01539   &   5         \\
41348.300  $\pm$ 0.01     &     1070.0  &      $-$0.00892   &   6         \\
41587.445  $\pm$ 0.01     &     1156.0  &      $-$0.00537   &   7         \\
42071.284  $\pm$ 0.01     &     1330.0  &      $-$0.01070   &   8         \\
42289.565  $\pm$ 0.01     &     1408.5  &      $-$0.01579   &   9         \\
43453.3049 $\pm$ 0.005    &     1827.0  &      $-$0.00492   &  10         \\
44166.546  $\pm$ 0.01     &     2083.5  &      $-$0.01709   &   5         \\
46762.359  $\pm$ 0.01     &     3017.0  &      $-$0.00110   &   5         \\
46773.496  $\pm$ 0.01     &     3021.0  &         0.01305   &   5         \\
47471.435  $\pm$ 0.01     &     3272.0  &      $-$0.00730   &  11         \\
47535.380  $\pm$ 0.01     &     3295.0  &      $-$0.01873   &  12         \\
48233.375  $\pm$ 0.005    &     3546.0  &         0.01692   &  13         \\
48561.483  $\pm$ 0.005    &     3664.0  &         0.00061   &   5         \\
48625.4378 $\pm$ 0.005    &     3687.0  &      $-$0.00103   &  14         \\
52279.2978 $\pm$ 0.0032   &     5001.0  &         0.00009   &  15         \\
52949.4487 $\pm$ 0.0025   &     5242.0  &      $-$0.00121   &  15         \\
54787.49993 $\pm$ 0.00015 &      5903.0 &         0.00014   &  16         \\
55157.33430 $\pm$ 0.00026 &      6036.0 &      $-$0.00044   &  16         \\
\hline \end{tabular} \end{center} \end{table}

To provide individual $UBV$ colours and an independent check on the WASP data we acquired the photoelectric $UBV$ light curves of XY\,Cet obtained by \citet{SrivastavaPadalia75apss}. The individual data are unfortunately not tabulated in that work, but are plotted as a function of orbital phase (their Fig.\,1). We have manually extracted the data from this plot by displaying it on a computer screen and mouse-clicking on each datapoint. This procedure was not straightforward because the model light curve is plotted as filled circles on the same axes as the individual observations are plotted as open circles. This has the twin effects of improving the overall appearance of the light curves without having to increase the information content of the data, and of obscuring some observations with the symbols representing the model light curve. It should also be borne in mind that any imperfections in defining the axis positions will cause the datapoints to be systematically shifted or stretched in phase or magnitude. A final caveat is that the values of the datapoints will be affected by discretisation due to the finite number of pixels on the computer screen, but this effect should be small (the plotting symbols were roughly five pixels wide). This procedure resulted in light curves containing 270 $U$, 263 $B$ and 280 $V$ datapoints; these numbers are rather smaller than the original work (335, 343 and 343, respectively) due to the difficulty of discerning overlapping plot symbols. \reff{The original data were made available to \citet{Okazaki89apss}, who published normal points which can be used as an alternative to the digitised version obtained above.}


\section{Period determination}                                                                                                      \label{sec:period}

\begin{figure*}\includegraphics[width=\textwidth,angle=0]{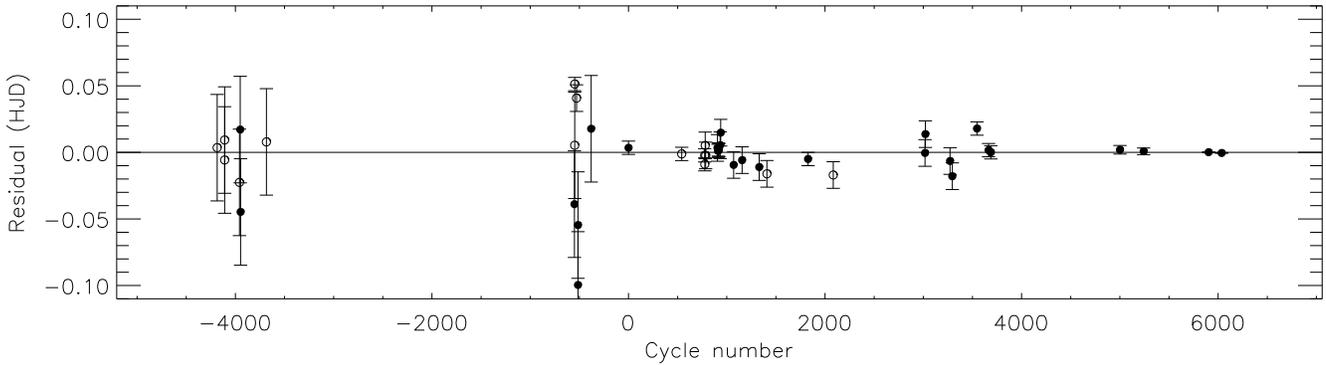}\\
\caption{\label{fig:periodresids} Residuals of the times of minimum
light around the ephemeris which best fits the observed times of
minima. The filled circles refer to primary minima and the open
circles to secondary minima.} \end{figure*}

All available times of minimum light of XY\,Cet were collected from the literature, and the orbital ephemeris given in the {\it General Catalogue of Variable Stars}\footnote{\tt http://www.sai.msu.su/groups/cluster/gcvs/gcvs/} was used to determine the preliminary cycle number of each minimum. Very few of the minimum timings are accompanied by an errorbar, so uncertainties were assigned based on the size of the residuals for each type of observing equipment used. Two additional times of minimum were calculated from the combined WASP data, separated into the 2008 and 2009 seasons, in the course of the analysis presented in Sect.\,\ref{sec:lc}.

A straight line was fitted to the resulting cycle numbers and times of minima (Table\,\ref{tab:minima}) by $\chi^2$ minimisation. The resulting ephemeris is:
\[ {\rm Min\,I} = {\rm HJD}\ 2438372.9455 (16) + 2.78071392 (28) \times E \]
where the bracketed quantity is the uncertainty in the final digit of the preceding number. All uncertainties quoted in this work are standard errors unless otherwise stated. The residuals of the fit are plotted in Fig.\,\ref{fig:periodresids} and give no indication of any form of period change. The above ephemeris is in good agreement with that given in the {\it Eclipsing Binaries Minima Database}\footnote{\tt http://www.as.ap.krakow.pl/} \citep{Kreiner++01book}, and shows no substantive evidence for departures from a constant period. There is also no evidence for the secondary minima being shifted from phase 0.5, implying that eccentricity is negligible.


\section{Radial velocity analysis}                                                                                                    \label{sec:spec}

\begin{table*} \centering
\caption{\label{tab:k1k2} Velocity amplitudes for the components of XY\,Cet found from
\reff{four} different analysis methods. The probable errors given by \citet{Popper71apj}
have been multiplied by 1.48 to transform them into standard errors.}
\begin{tabular}{llr@{\,$\pm$\,}lr@{\,$\pm$\,}lr@{\,$\pm$\,}lr@{\,$\pm$\,}ll} \hline
Analysis type & Template spectrum & \multicolumn{4}{c}{Uncorrected} & \multicolumn{4}{c}{Corrected} & Light ratio \\
\ & & \mc{$K_{\rm A}$ (\kms)} & \mc{$K_{\rm B}$ (\kms)} & \mc{$K_{\rm A}$ (\kms)} & \mc{$K_{\rm B}$ (\kms)} \\ \hline
{\sc onecor} & HD 24740 & 108.23 & 0.27 & 118.65 & 0.45 & 108.24 & 0.23 & 118.65 & 0.28 \\
{\sc onecor} & HD 905   & 108.12 & 0.26 & 118.66 & 0.48 & 108.10 & 0.21 & 118.65 & 0.21 \\
{\sc onecor} & HD 32115 & 108.10 & 0.28 & 118.72 & 0.43 & 108.15 & 0.22 & 118.72 & 0.22 \\
{\sc onecor} & HD 37594 & 108.02 & 0.26 & 118.74 & 0.40 & 107.92 & 0.21 & 118.74 & 0.20 \\
{\sc onecor} & HD 39945 & 107.86 & 0.26 & 118.66 & 0.44 & 107.86 & 0.23 & 118.65 & 0.21 \\[2pt]
{\sc todcor} & HD 24740 & 108.42 & 0.22 & 118.89 & 0.34 & 108.42 & 0.22 & 118.89 & 0.22 & $0.72 \pm 0.11$ \\
{\sc todcor} & HD 905   & 108.32 & 0.21 & 118.95 & 0.34 & 108.13 & 0.09 & 118.89 & 0.20 & $0.69 \pm 0.13$ \\
{\sc todcor} & HD 32115 & 108.33 & 0.23 & 119.12 & 0.33 & 108.31 & 0.22 & 119.14 & 0.26 & $0.73 \pm 0.10$ \\
{\sc todcor} & HD 37594 & 108.12 & 0.22 & 119.24 & 0.30 & 108.12 & 0.22 & 119.22 & 0.24 & $0.72 \pm 0.10$ \\
{\sc todcor} & HD 39945 & 108.02 & 0.23 & 118.94 & 0.33 & 108.03 & 0.21 & 118.94 & 0.25 & $0.70 \pm 0.11$ \\[2pt]
{\sc FDbinary} & \ \ -- & 108.06 & 0.30 & 119.22 & 0.35 \\
\citet{Popper71apj} &   & 109.6  & 0.6  & 117.6  & 1.0  \\
\hline \end{tabular} \end{table*}

Our time-resolved spectroscopic dataset consists of 38 observations covering the wavelength range 4230--4500\,\AA, of which six were taken during secondary eclipse when the two stars have very similar velocities. We have measured radial velocities (RVs) from the remaining 32 spectra, concentrating on the 4360--4490\,\AA\ wavelength range which contains a multitude of spectral lines but avoids the very broad H$\gamma$ feature. We have considered three different methods of measuring the velocity amplitudes of the component stars of XY\,Cet, and this redundancy allows consistency checks and the assignment of robust measurement uncertainties.

A large number of standard stars were observed using the same observational setup as for our target star. Inspection of these yielded five which have a similar appearance to the spectra of the components of XY\,Cet: HD\,39945 (spectral type A5\,V), HD\,32115 (A8\,IV), HD\,37594 (A8\,Vs), HD\,905 (F0\,IV) and HD\,24740 (F2\,IV). These will be used as template spectra in the analyses below.

\subsection{{\sc onecor}: standard cross-correlation}

Numerical cross-correlation \citep{Simkin74aa,TonryDavis79aa} is a standard approach for measuring RVs from the spectra of celestial objects. We used our own implementation of this method ({\sc onecor}; \citealt{MeClausen07aa}) after binning all spectra onto a common logarithmic wavelength scale. The cross-correlation functions (CCFs) were interactively assigned weights based on two factors: signal to noise and the velocity separation of the components of XY\,Cet. These weights were fixed for all subsequent analyses, and we have verified that their precise values do not have a significant effect on the resulting RVs.

Each template spectrum was cross-correlated against the spectra of XY\,Cet, and the positions of the two peaks were measured using quadratic interpolation. The resulting RVs were fitted with spectroscopic orbits using the {\sc sbop}\footnote{Spectroscopic Binary Orbit Program, written by P.\ B.\ Etzel, \\ {\tt http://mintaka.sdsu.edu/faculty/etzel/}} code. Orbits were fitted for the two stars separately (see \citealt{Me++04mn} and \citealt{PopperHill91aj}). Fits including orbital eccentricity yielded values which were small and not significantly different from zero, as expected from the period study (Sect.\,\ref{sec:period}), so our final results were calculated with eccentricity fixed at zero. The outcome of this analysis was velocity amplitude measurements for the two \reff{components} of XY\,Cet ($K_{\rm A}$ and $K_{\rm B}$) and for each of the five template spectra (Table\,\ref{tab:k1k2}).

RVs found from cross-correlation analyses are known to show slight biases due to effects such as line blending \citep{PetrieAndrews66aa} and individual spectral lines being Doppler-shifted into or out of the considered wavelength range \citep[e.g.][]{Torres+97aj,Clausen+08aa}. These biases can be measured and thus removed by constructing synthetic composite spectra with known RVs and measuring them in the same way as the observed spectra. We have performed this analysis using our five observed template spectra and the method discussed by \citet{MeClausen07aa}. We find that removing these biases from the measured RVs results in velocity amplitudes whose values are almost unchanged but whose uncertainties are noticeably lower; the results are given in Table\,\ref{tab:k1k2}.

\subsection{{\sc todcor}: two-dimensional cross-correlation}

The {\sc todcor} algorithm, introduced by \citet{ZuckerMazeh94apj} calculates a two-dimensional CCF for a double-lined spectrum using two template spectra. This approach is aimed at avoiding line blending when the two target stars have quite different spectral characteristics, although \citet{MeClausen07aa} found that it is no better than {\sc onecor} for spectroscopic binaries containing two similar stars. This is the case with XY\,Cet, but the relatively low rotational velocities of its components mean that line blending is not a significant problem.

We used our own implementation of {\sc todcor} \citep{Me++04mn} and the same method and template stars as for {\sc onecor} in deriving $K_{\rm A}$ and $K_{\rm B}$. This process included the measurement and removal of RV biases, using synthetic spectra constructed from the observed template spectra. In each case we used the same template spectrum for both target stars, as the two components of XY\,Cet have similar spectral characteristics.

One advantage of {\sc todcor} over {\sc onecor} is that the optimal light ratio between the two target stars can be calculated as part of the algorithm. These light ratios are given in Table\,\ref{tab:k1k2} but come with a health warning. Both components of XY\,Cet show the Am phenomenon, which means their spectral line strengths are peculiar and do not reflect their overall chemical composition. The light ratios from {\sc todcor} strictly refer to the overall line strengths rather than the integrated light of each star. We have also not adjusted them for the difference in equivalent widths caused by the different effective temperatures of the two stars. They therefore cannot be used as observational constraints, but should be good enough to act as checks on the results of the light curve analysis below.

\begin{figure} \includegraphics[width=0.48\textwidth,angle=0]{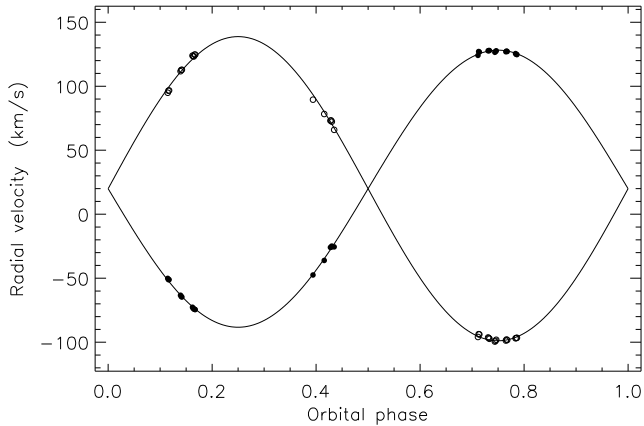}
\caption{\label{fig:uncor} RVs and best-fitting spectroscopic orbit measured
using {\sc todcor} and HD\,905 as a template star. No velocity corrections
have been made. Filled circles represent RVs for star A and open circles
those for star B.} \end{figure}

\begin{figure} \includegraphics[width=0.48\textwidth,angle=0]{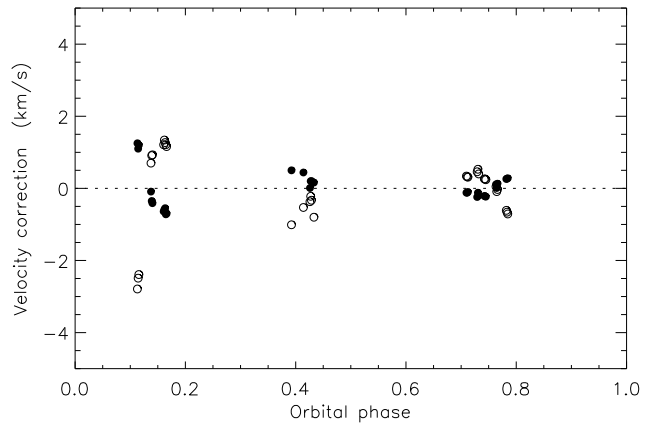}
\caption{\label{fig:rvcor} RV corrections measured for the {\sc todcor}
analysis with HD\,905 as a template star. Filled circles represent
the corrections for star A and open circles those for star B.} \end{figure}

\begin{figure} \includegraphics[width=0.48\textwidth,angle=0]{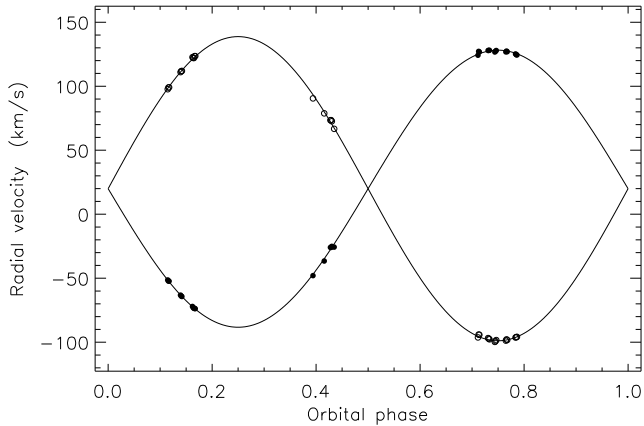}
\caption{\label{fig:cor} RVs and best-fitting spectroscopic orbit measured
using {\sc todcor} and HD\,905 as a template star, including velocity
corrections. Filled circles represent RVs for star A and open circles
those for star B.} \end{figure}

The {\sc todcor} results are in general slightly better than the {\sc onecor} results: the $K_{\rm A}$ and $K_{\rm B}$ values are very similar but the bias correction has a smaller effect (Table\,\ref{tab:k1k2}). For our final results we adopt the values calculated using HD\,905 as a template star. The corresponding RVs are given in Table\,\ref{tab:rvs}. The uncorrected spectoscopic orbits are plotted in Fig.\,\ref{fig:uncor}, the RV corrections are shown in Fig.\,\ref{fig:rvcor} and the final spectroscopic orbits are displayaed in Fig.\,\ref{fig:cor}.

\begin{table} \centering \caption{\label{tab:rvs} Corrected RVs from a
{\sc todcor} analysis of the spectra of XY\,Cet using as templates a
spectrum of HD\,905. The columns marked $O-C$ give the observed minus
calculated velocity residuals.}
\begin{tabular}{lrrrrr} \hline
HJD $-$ & Weight & \mc{Radial velocities} & \mc{$O-C$} \\
2\,400\,000&&star A&star B&\multicolumn{1}{c}{A}&\multicolumn{1}{c}{B}\\
\hline
52569.58310 & 2 & $-$47.97 &    90.50 & $-$0.95 & $-$2.90 \\
52569.64324 & 1 & $-$36.55 &    78.81 & $-$1.67 & $-$1.27 \\
52569.69624 & 1 & $-$25.67 &    66.71 & $-$2.34 & $-$0.70 \\
52570.46545 & 2 &   124.21 & $-$96.25 & $-$0.63 & $-$1.13 \\
52570.46911 & 2 &   127.24 & $-$94.01 &    2.19 &    1.34 \\
52570.47278 & 2 &   126.82 & $-$94.20 &    1.56 &    1.38 \\
52570.51954 & 2 &   127.92 & $-$96.77 &    0.69 &    0.97 \\
52570.52321 & 2 &   128.00 & $-$97.25 &    0.66 &    0.61 \\
52570.52686 & 2 &   128.08 & $-$97.72 &    0.65 &    0.24 \\
52570.55581 & 2 &   126.73 & $-$99.75 & $-$1.21 & $-$1.23 \\
52570.55948 & 2 &   127.34 & $-$99.15 & $-$0.63 & $-$0.60 \\
52570.56315 & 2 &   127.91 & $-$98.32 & $-$0.09 &    0.26 \\
52570.61455 & 2 &   126.81 & $-$98.74 & $-$0.74 & $-$0.65 \\
52570.61821 & 2 &   127.46 & $-$97.79 &    0.00 &    0.20 \\
52570.62188 & 2 &   126.99 & $-$98.25 & $-$0.38 & $-$0.36 \\
52570.66646 & 2 &   125.25 & $-$96.31 & $-$0.37 & $-$0.34 \\
52570.67013 & 2 &   124.63 & $-$96.06 & $-$0.80 & $-$0.30 \\
52570.67380 & 2 &   124.42 & $-$95.81 & $-$0.81 & $-$0.27 \\
52571.58690 & 2 & $-$51.35 &    97.69 &    0.17 & $-$0.65 \\
52571.59056 & 2 & $-$52.21 &    99.01 & $-$0.02 & $-$0.07 \\
52571.59423 & 2 & $-$52.31 &    99.11 &    0.54 & $-$0.70 \\
52571.65538 & 2 & $-$63.19 &   111.02 &    0.01 & $-$0.14 \\
52571.65904 & 2 & $-$63.94 &   111.77 & $-$0.17 & $-$0.01 \\
52571.66270 & 2 & $-$64.29 &   111.95 &    0.05 & $-$0.45 \\
52571.71929 & 2 & $-$72.04 &   122.68 &    0.28 &    1.52 \\
52571.72295 & 2 & $-$72.89 &   122.14 & $-$0.11 &    0.47 \\
52571.72663 & 2 & $-$73.54 &   122.01 & $-$0.29 & $-$0.17 \\
52571.73029 & 2 & $-$73.57 &   123.49 &    0.13 &    0.82 \\
52571.73395 & 2 & $-$73.79 &   123.54 &    0.35 &    0.38 \\
52572.45708 & 1 & $-$25.92 &    73.45 &    1.83 &    1.20 \\
52572.46075 & 1 & $-$25.91 &    73.50 &    1.03 &    2.13 \\
52572.46442 & 1 & $-$25.10 &    72.65 &    1.03 &    2.17 \\
\hline \end{tabular} \end{table}

\subsection{Final spectroscopic results}

\begin{table} \begin{center} \caption{\label{tab:orbit}
Final spectroscopic orbital parameters for XY\,Cet.}
\begin{tabular}{l r@{\,$\pm$\,}l r@{\,$\pm$\,}l} \hline
Parameter                     &  \mc{Primary}  & \mc{Secondary} \\ \hline
Velocity amplitude $K$ (\kms) &  108.3 & 0.4   &  118.9 & 0.5   \\
Systemic velocity (\kms)      & $-$6.0 & 1.3   & $-$5.7 & 1.3   \\
Mass ratio $q$                & \multicolumn{4}{c}{$0.9108 \pm 0.0051$}\\
$a \sin i$ (\Rsun)            & \multicolumn{4}{c}{$12.482 \pm 0.035$} \\
$M \sin^3 i$ (\Msun)          &  1.768 & 0.016 &  1.611 & 0.014 \\
\hline \end{tabular}\end{center} \end{table}

The full set of $K_{\rm A}$ and $K_{\rm B}$ measurements is collected in Table\,\ref{tab:k1k2}, and includes values measured by spectral disentangling in Sect.\,\ref{sec:spd}. The values demonstrate a good agreement between different analysis methods and different template spectra. We also find an acceptable agreement with the results of \citet{Popper71apj}. For our final values we adopt the $K_{\rm A}$ and $K_{\rm B}$ found via {\sc todcor} with HD\,905 as a template, but we increase the uncertainties in these quantities to account for the variations in velocity amplitudes throughout Table\,\ref{tab:k1k2}. The systemic velocities are measured relative to HD\,905, for which we adopt a systemic velocity of $-25.8 \pm 1.3$\kms\ \citep{Nordstrom+04aa}. Table\,\ref{tab:orbit} gives final orbital parameters.


\section{Light curve analysis}                                                                                                          \label{sec:lc}

\begin{table} \begin{center}
\caption{\label{tab:lcfit} Model parameters of the WASP light curve of XY\,Cet.}
\begin{tabular}{l l r@{\,$\pm$\,}l} \hline
Parameter                        &                             &    \mc{Value}      \\
\hline
Sum of the relative radii        & $r_{\rm A}+r_{\rm B}$       &  0.2918 & 0.0011   \\
Ratio of the radii               & $k$                         &   0.947 & 0.032    \\
Central surface brightness ratio & $J$                         &   0.806 & 0.043    \\
Orbital inclination (\degr)      & $i$                         &   87.66 & 0.12     \\
Linear LD coefficient star\,A    & $u_{\rm A}$                 &    0.49 & 0.10     \\
Linear LD coefficient star\,B    & $u_{\rm B}$                 &    0.40 & 0.12     \\
$e \cos \omega$                  &                             & 0.00021 & 0.00022  \\[2pt]
Relative radius of star\,A       & $r_{\rm A}$                 &  0.1499 & 0.0028   \\
Relative radius of star\,B       & $r_{\rm B}$                 &  0.1419 & 0.0023   \\
Light ratio                      & $\ell_{\rm B}/\ell_{\rm A}$ &   0.752 & 0.061    \\
\hline \end{tabular} \end{center} \end{table}

\begin{figure*} \includegraphics[width=\textwidth,angle=0]{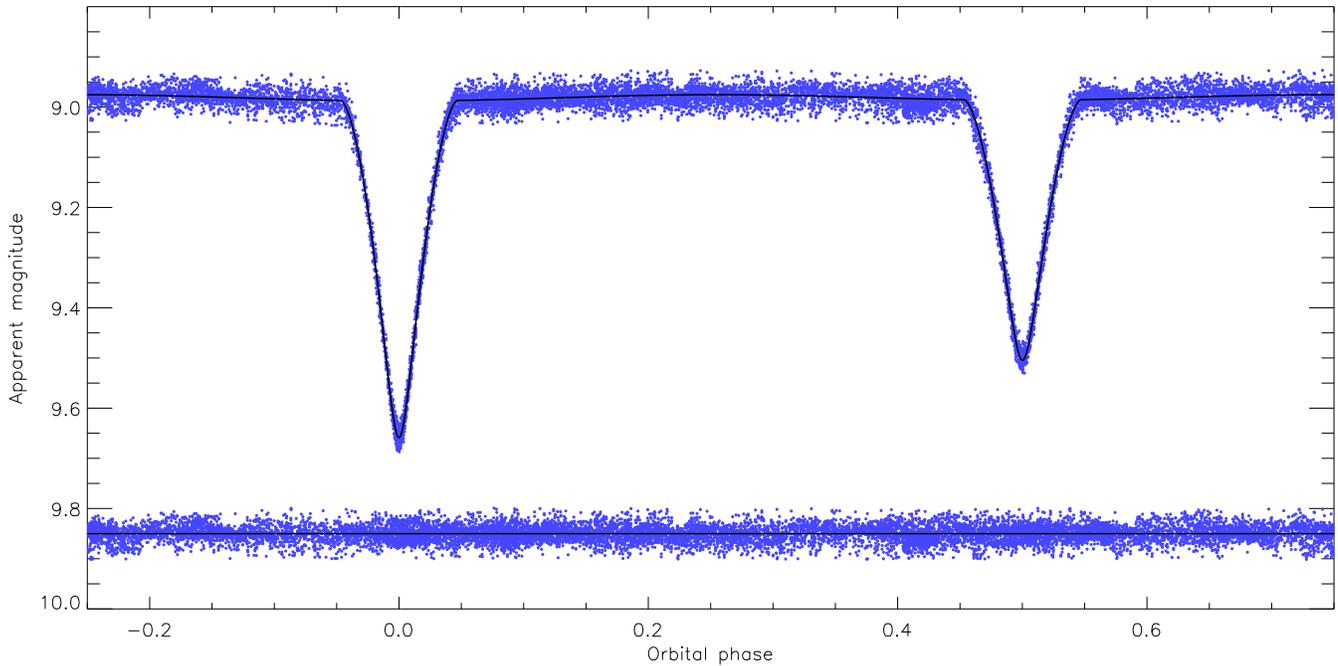}
\caption{\label{fig:lcwasp} Plot of the WASP light curve of XY\,Cet (filled
grey circles) compared to the best-fit model (black line). The residuals of the
fit are plotted at the base of the figure with an offset from zero.} \end{figure*}

\begin{table} \begin{center}
\caption{\label{tab:ubv} Combined and individual $UBV$ and $uvby$ indices
for XY\,Cet. The combined $UBV$ and $uvby$ measurements are taken from
\citet{PopperDumont77} and \citet{HilditchHill75mmras}, respectively.}
\begin{tabular}{l r@{\,$\pm$\,}l r@{\,$\pm$\,}l r@{\,$\pm$\,}l} \hline
Parameter     & \mc{Combined} &   \mc{Star A}  &   \mc{Star B}   \\
\hline
$V$           &  8.78 & 0.07  &    9.38 & 0.07 &    9.70 & 0.07  \\
$B-V$         &  0.26 & 0.01  &    0.21 & 0.02 &    0.33 & 0.03  \\
$U-B$         &  0.14 & 0.01  &    0.16 & 0.02 &    0.11 & 0.03  \\[3pt]
$V$           &  8.74 & 0.05  &    9.27 & 0.06 &    9.77 & 0.07  \\
$b-y$         &  0.16 & 0.01  &    0.14 & 0.04 &    0.18 & 0.06  \\
$m_1$         &  0.23 & 0.01  &    0.27 & 0.05 &    0.17 & 0.08  \\
$c_1$         &  0.82 & 0.02  &    0.87 & 0.04 &    0.76 & 0.06  \\[3pt]
\Teff\ (K)    &  7750 & 240   &    7960 & 560  &    7570 & 730   \\
Spectral type &   \mc{ }      &     \mc{A8}    &     \mc{F0}     \\
\hline \end{tabular} \end{center} \end{table}

\begin{figure} \includegraphics[width=\columnwidth,angle=0]{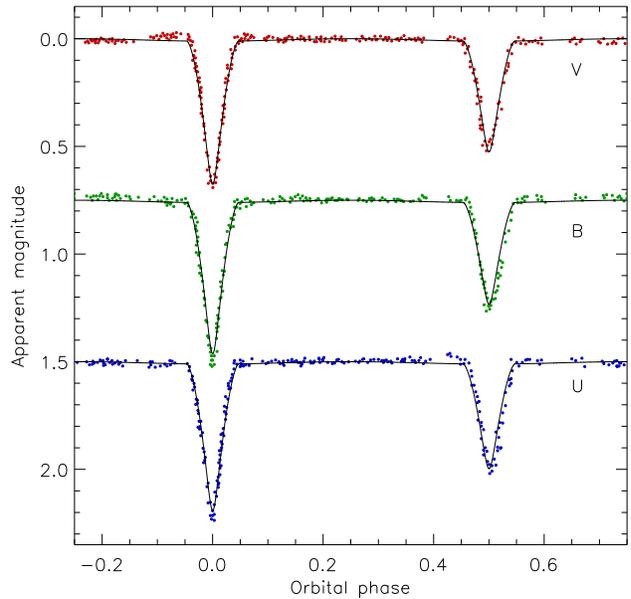}
\caption{\label{fig:lcsriv} Plot of the $UBV$ light curves of XY\,Cet from
\citet{SrivastavaPadalia75apss} compared to {\sc jktebop} models in which
only the surface brightness ratio is an adjusted parameter.} \end{figure}

We have modelled the WASP light curve of XY\,Cet using the {\sc jktebop} code\footnote{{\sc jktebop} is written in {\sc fortran77} and the source code is available at {\tt http://www.astro.keele.ac.uk/$\sim$jkt/codes.html}} code \citep{Me++04mn,Me++04mn2}, which is based on the {\sc ebop} program \citep{PopperEtzel81aj,Etzel81conf,NelsonDavis72apj}. The fractional radii of the two stars ($r_{\rm A} = \frac{R_{\rm A}}{a}$ and $r_{\rm B} = \frac{R_{\rm B}}{a}$ where $a$ is the orbital semimajor axis) are reparameterised as their sum and ratio: $r_{\rm A}+r_{\rm B}$ and $k=\frac{r_{\rm B}}{r_{\rm A}}$ because these are more directly related to the morphology of the light curve. Aside from $r_{\rm A}+r_{\rm B}$ and $k$ we included the orbital inclination ($i$), and the ratio of the central surface brightnesses of the stars ($J$) as fitted parameters. The gravity darkening exponents were set to 1.0 and the mass ratio (which has a negligible effect on the photometric solution) was fixed to 0.91. Attempts to fit for third light turned up a small and sub-zero value, so this parameter was fixed at zero for the final solution.

For limb darkening (LD) we used the linear law \citep[e.g.][]{Me08mn} and the good quality of the light curve meant we were able to include the LD coefficient for both stars as fitted parameters. An alternative solution using the quadratic law, with the linear coefficient for star included as a fitted parameter, produced results which were practically identical to those obtained via the linear law. The resulting linear LD coefficients, $u_{\rm A} = 0.49 \pm 0.10$ and $u_{\rm B} = 0.40 \pm 0.12$, are a little lower than the theoretically expected value of 0.58 for the $g$$+$$r$ passband \citep{Claret04aa2}. This cannot be attributed to the metallic nature of the stars, as coefficients for $[\frac{\rm M}{\rm H}] = +0.5$ are very similar to those for solar metallicity.

The orbital period was fixed to the value found in Sect.\,\ref{sec:period} but the time of primary mid-eclipse was allowed to vary. A check for orbital eccentricity returned a positive result, so the quantity $e\cos\omega$ (where $e$ is eccentricity and $\omega$ is periastron longitude) was also included as an adjusted parameter. The final value for this quantity, $e\cos\omega = 0.00021 \pm 0.00020$, is not significant and is too diminutive to affect either the photometric or spectroscopic results. The final solution of the WASP light curve is given in Table\,\ref{tab:lcfit} and plotted in Fig.\,\ref{fig:lcwasp}.

To find the 1$\sigma$ uncertainties in the measured parameters, we have run both 10\,000 Monte Carlo simulations \citep{Me+04mn3} and a residual-permutation analysis \citep{Jenkins++02apj}. The latter approach is able to account for any correlated errors (red noise) in the WASP data. We find, as expected, that the residual-permutation analysis returns larger error estimates than the Monte Carlo simulations. The residual-permutation error estimates are therefore adopted as the final errorbars in Table\,\ref{tab:lcfit}.

\subsection{Checks, comparisons and colour indices}

We checked the {\sc jktebop} solution of the WASP data using the 2003 version of the Wilson-Devinney code \citep{WilsonDevinney71apj}. The data were phase-binned into 140 points with the eclipse phases sampled more finely than the other phases, then fitted using the {\sc jktwd} wrapper to the Wilson-Devinney code \citep{Me+11mn}. The resulting parameter values are consistent with {\sc jktebop} to within the errors, indicating that both solutions are reliable.

The \citet{SrivastavaPadalia75apss} light curves were modelled using {\sc jktebop} with the geometrical parameters ($r_{\rm A}$, $r_{\rm B}$ and $i$) fixed, the LD coefficients set to theoretically expected values \citep{Vanhamme93aj,Claret00aa,ClaretHauschildt03aa}, and only $J$ and the out-of-eclipse brightness adjusted. A good fit was found, corroborating the WASP light curve solution, and was used to obtain the individual $UBV$ magnitudes and colours of the two stars (Table\,\ref{tab:ubv}). These should be treated with caution due to the modest quality of the data and the way in which they were obtained for this study. A plot of the \citet{SrivastavaPadalia75apss} light curves and the {\sc jktebop} constrained fit is shown in Fig.\,\ref{fig:lcsriv}. \reff{Analysis of the normal points published by \citet{Okazaki89apss} yielded values which were very similar, indicating that our digitised version of the \citet{SrivastavaPadalia75apss} data has some reliability.}

Str\"omgren $uvby$ photometry \citep{Stromgren66aarv} of XY\,Cet was obtained at five different orbital phases by \citet{HilditchHill75mmras}. One of these is during primary eclipse, which allows individual Str\"omgren indices to be obtained for the two stars. These Str\"omgren index measurements were converted into $uvby$ light curves and fitted using {\sc jktebop}, with all parameters except $J$ and the out-of-eclipse brightness fixed. From the resulting light ratios and the combined Str\"omgren indices we have obtained the indices for the individual stars (Table\,\ref{tab:ubv}). Using the $uvby$ grids of \citet{SmalleyKupka97aa} we find effective temperatures for the two stars of ${\Teff}_{\rm A} = 7960$\,K and ${\Teff}_{\rm B} = 7570$\,K. From the calibration of $b-y$ versus spectral type \citep{Stromgren63qjras} we find types of A8 and F0.

As a final investigation of the light curve we calculated a periodogram of the residuals of the best {\sc jktebop} fit using the {\sc Period04} package \citep{LenzBreger04iaus}. There is a peak of height 5\,mmag at 1 cycle per day, and lower peaks at the multiples of this frequency, which are certainly of instrumental origin. There are no other clear periodicities in the frequency range of $\gamma$\,Doradus pulsations or $\delta$\,Scuti pulsations, to limiting amplitudes of 2\,mmag and 1\,mmag respectively. The components of XY\,Cet are therefore not found to be pulsating stars.


\section{Atmospheric parameters from spectral disentangling}                                                           \label{sec:atm} \label{sec:spd}

Our spectroscopic dataset is well suited to a spectral disentangling analysis, in which the individual spectra and velocity amplitudes of the two stars are extracted from a set of phase-resolved spectra \citep{SimonSturm94aa,PavlovskiHensberge09xxx}. The resulting disentangled spectra contain the total signal of the input spectra and are well suited to derivation of the atmospheric parameters and chemical abundances of the stars \citep[e.g.][]{PavlovskiMe09mn,Pavlovski+09mn}.

\begin{figure} \includegraphics[width=\columnwidth,angle=0]{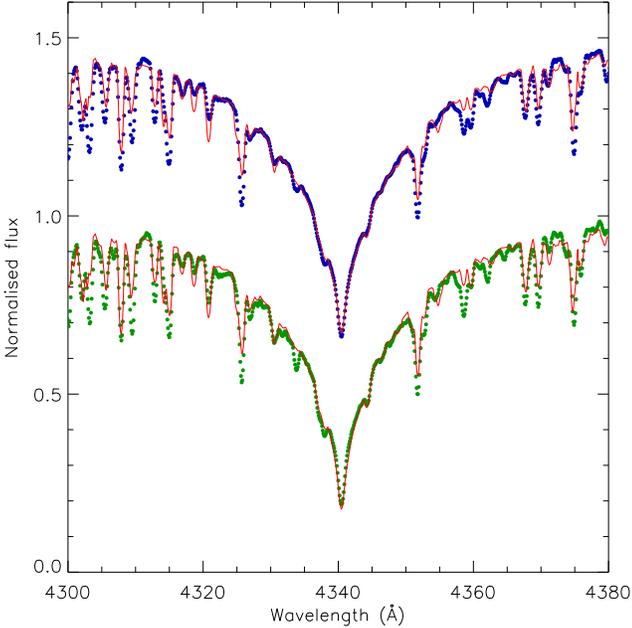}
\caption{\label{fig:hgamma} Plot of the disentangled spectra of star\,A (blue points,
offset by $+0.5$) and star\,B (green points) around the H$\gamma$ line. The best-fitting
synthetic spectra for the two stars are shown with thin red lines.} \end{figure}

We carefully normalised and velocity-binned the spectra and disentangled them using the {\sc FDbinary}\footnote{\tt http://sail.zpf.fer.hr/fdbinary/} package \citep{Ilijic+04aspc}. This analysis was initially performed both with and without the six spectra taken during secondary eclipse. We found that the \reff{former} approach resulted in much more tightly constrained solutions, as these spectra contain information on how much light is lost from the secondary star at two orbital phases during secondary eclipse. The velocity amplitudes we find are included in Table\,\ref{tab:k1k2}.

We measured the projected rotational velocities of the two stars by fitting the disentangled spectra and the {\sc uclsyn} program \citep{Smalley++01,Smith92phd}, finding $v_{\rm A} \sin i = 34.4 \pm 0.4$\kms\ and $v_{\rm B} \sin i = 34.1 \pm 0.5$\kms. These are close to the equatorial rotational velocity values, under the assumption that the rotational axes of the stars are aligned with their orbital axis. The rotational velocity of star\,A is therefore synchronous (Table\,\ref{tab:absdim}). Star\,B is rotating slightly faster than the synchronous velocity, although this is only significant at the $2.5\sigma$ level.

The disentangled spectra of the stars contain their H$\gamma$ lines, which are a useful indicator of \Teff\ for A and F stars but suffer from a degeneracy between \Teff\ and surface gravity \citep{Smalley05msais,Smalley05msais2}. We have pursued two methods of obtaining \Teff\ from the spectra, in both cases fixing the surface gravity (\logg) values to those found from the masses and radii of the stars (see Sect.\,\ref{sec:absdim}), the relative light contributions of the stars to the value obtained from the light curve analysis (Sect.\,\ref{sec:lc}) and the $v\sin i$ values to the numbers above.

\begin{figure} \includegraphics[width=\columnwidth,angle=0]{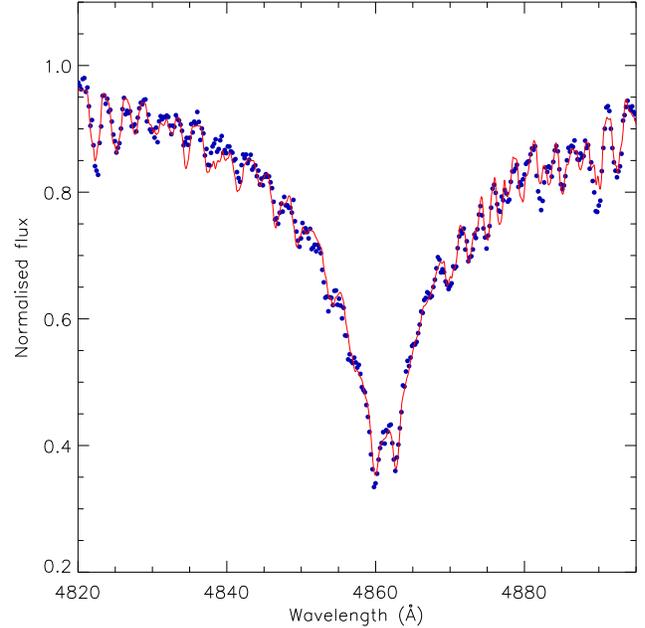}
\caption{\label{fig:hbeta} Comparison of the spectrum around H$\beta$
observed at orbital phase 0.143 (blue points binned by two for clarity)
to a composite synthetic spectrum generated for the atmospheric parameters
we found for the components of XY\,Cet and their expected radial velocities
at this orbital phase (thin red line).} \end{figure}

Method 1 is a one-dimensional interpolation in \Teff\ within a grid of LTE synthetic spectra calculated using the {\sc spectrum}\footnote{\tt http://www.phys.appstate.edu/spectrum/spectrum.html} code \citep{GrayCorbally94aj} and {\sc atlas9} model atmospheres \citep{Kurucz93}. A grid of spectra centred on H$\gamma$ was constructed, covering \Teff s from 7000 to 10\,000\,K in steps of 200\,K. Residuals were calculated between the observed and synthetic spectra, in carefully-selected spectral intervals which are free of metal lines. This approach yielded ${\Teff}_{\rm,A} = 7870 \pm 130$\,K and ${\Teff}_{\rm,B} = 7620 \pm 170$\,K.

\begin{figure*} \includegraphics[width=\textwidth,angle=0]{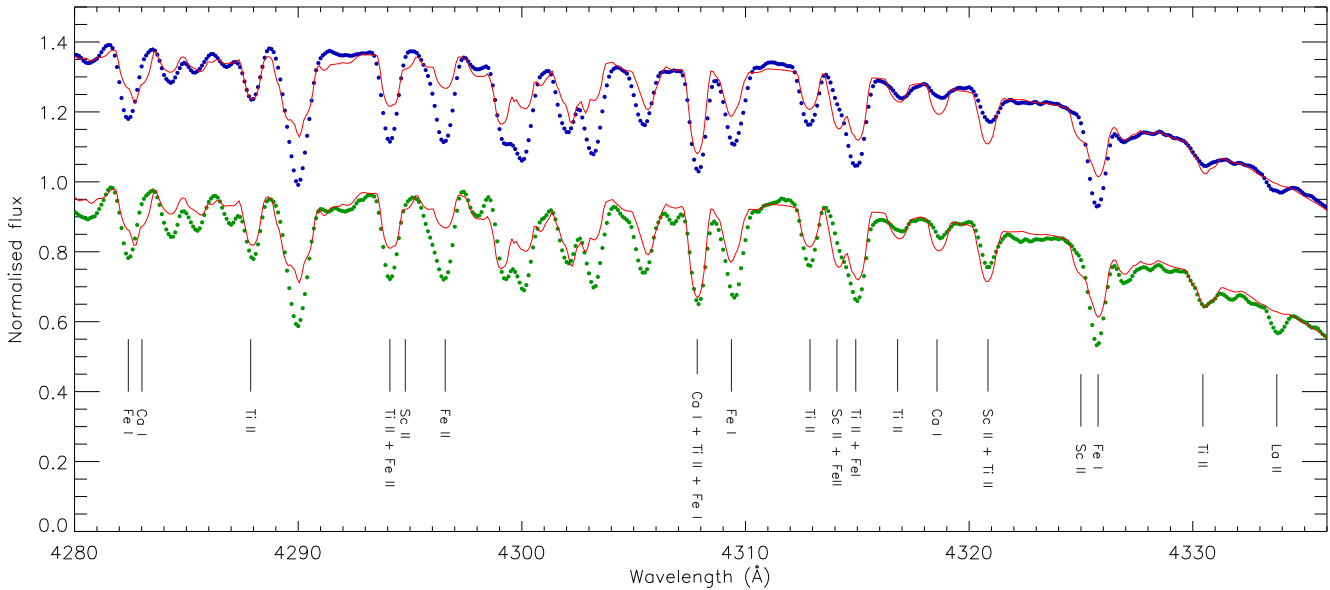}
\caption{\label{fig:4310} The observed and disentangled spectra of XY\,Cet
in the 4300--4340\,\AA\ wavelength region. Star\,A is plotted using blue
points offset by $+0.4$ and star\,B is plotted using green points. Synthetic
spectra calculated for the appropriate \Teff s, \logg s and \Vsini s and a
solar chemical composition are shown by red lines. Some absorption lines are
labelled with the species which causes them.} \end{figure*}

Method 2 utilised the {\sc genfitt} code \citep{Tamajo++10aa}, which fits synthetic spectra to hydrogen line profiles using a genetic algorithm. This method yielded ${\Teff}_{\rm,A} = 7870 \pm 115$\,K and ${\Teff}_{\rm,B} = 7650 \pm 125$\,K, in good agreement with method 1. A comparison between the observed and calculated H$\gamma$ line profiles is shown in Fig.\,\ref{fig:hgamma}. An attempt to fit for \logg\ as well as \Teff\ returned less precise and somewhat different values, underlining a big advantage of dEBs that \logg\ can be fixed in such analyses. We also obtained a solution with the light ratio fitted for and found $0.72 \pm 0.08$, in agreement with the light curve results. The relatively large uncertainty arises because the H$\gamma$ profiles contain many metal lines which have to be masked out in the {\sc genfitt} analysis, resulting in a significant loss of information. Better results could be obtained for the H$\alpha$ or H$\beta$ profiles, as these have far fewer metal lines, or for stars which are hotter or metal-poor.

The \Teff s found above are in good agreement with those coming from the Str\"omgren indices (Table\,\ref{tab:ubv}). We therefore adopt $7870 \pm 115$\,K and $7620 \pm 125$\,K as our final values. The components of XY\,Cet turn out to be substantially cooler than \reff{was} previously thought. As a test we generated solar-abundance synthetic spectra covering the H$\beta$ region, and combined them using the appropriate light ratio and RVs for our single observed spectrum of H$\beta$. Fig.\,\ref{fig:hbeta} shows that there is an excellent agreement between the synthetic composite spectrum and our observations. Our data rule out the hotter \Teff s found in previous studies to high confidence.

\subsection{Chemical abundances}

A major motivation for the current study was the metallic-lined nature of the components of XY\,Cet, which makes their physical properties useful in investigating the Am phenomenon. The limited wavelength coverage of our spectroscopic data unfortunately precludes a detailed abundance analysis, but it is possible to obtain some qualitiative results by comparing the disentangled spectra to synthetic spectra calculated for a solar chemical composition. We adopted a microturburbulence velocity of 2\kms\ but our conclusions are robust against an increase to a value of 4\kms, as often found for Am stars \citep[e.g.][]{CoupryBurkhart92aas}. Fig.\,\ref{fig:4310} shows this comparison for the 4300--4340\,\AA\ wavelength region, with a selection of absorption lines labelled.

From Fig.\,\ref{fig:4310} it is obvious that most of the spectral lines are significantly enhanced compared to solar abundance, which is a common feature of Am stars. The prime hallmark of the Am phenomenon is that Ca and Sc are underabundant in the photosphere, causing their absorption lines to be unusually weak. Ca\,I lines can be found at 4283.011, 4318.562 and 4455.887 \AA, and Sc\,II lines at 4294.787, 4314.089, 4320.732, 4324.996 and 4400.389 \AA. These are uniformly weaker than solar, most obviously in the region of 4320\,\AA\ (Fig.\,\ref{fig:4310}), so we confirm that both components are Am stars.

Another intriguing facet of the Am phenomenon is a strong overabundance of rare earth elements \citep{Wolff83book}. We have therefore carefully combed the disentangled spectra to uncover absorption lines which are far stronger than expected for a solar composition. A number of good examples have been found, comprising La\,II 4333.753\,\AA\ (see Fig.\,\ref{fig:4310}), Ce\,II (4364.653 and 4449.30 \AA) and Nd\,II (4446.384\,\AA). Zr\,II (4403.349\,\AA) is also highly enhanced. These elements have overabundances of order 1\,dex in the photospheres of both components of XY\,Cet. This system is an excellent candidate for a detailed abundance analysis based on (\'echelle) spectra with a wide wavelength coverage and a high S/N ratio.


\section{The physical properties of XY\,Cet}                                                                                        \label{sec:absdim}

We have calculated the physical properties of the XY\,Cet system based on the results from the photometric and spectroscopic analyses above. This was done with the {\sc jktabsdim} code \citep{Me++05aa}, which calculates complete error budgets using a perturbation algorithm. The physical properties of XY\,Cet are given in Table\,\ref{tab:absdim} and show that we have measured the masses of the stars to accuracies of 1\% and the radii to accuracies of 2\%. Additional photometric observations could allow the radius measurements to be improved to the 1\% level or better. A spectroscopic light ratio would be beneficial, and could most usefully be obtained from disentangling spectra covering the H$\beta$ line.

\begin{table} \begin{center}
\caption{\label{tab:absdim} The physical properties of the XY\,Cet system.}
\begin{tabular}{l r@{\,$\pm$\,}l r@{\,$\pm$\,}l} \hline
Parameter                     &    \mc{Star A}   &    \mc{Star B}    \\ \hline
Orbital separation (\Rsun)    &\multicolumn{4}{c}{$12.493 \pm 0.035$}\\
Mass (\Msun)                  &  1.773  & 0.016  &  1.615  & 0.014   \\
Radius (\Rsun)                &  1.873  & 0.035  &  1.773  & 0.029   \\
$\log g$ [cm\,s$^{-2}$]       &  4.142  & 0.016  &  4.149  & 0.014   \\
\Vsync\ (\kms)                &  34.09  & 0.64   &  32.27  & 0.53    \\
\Vsini\ (\kms)                &  34.4   & 0.4    &  34.1   & 0.5     \\[3pt]
\Teff (K)                     &  7870   & 115    &  7620   & 125     \\
$\log(L/\Lsun)$ $*$           &  1.082  & 0.030  &  0.978  & 0.032   \\
\Mbol\ $*$                    &  2.045  & 0.076  &  2.304  & 0.080   \\
Distance ($V$-band, pc)       &  285    & 13     &  292    & 14      \\
Distance ($K$-band, pc)       &  \multicolumn{4}{c}{$279.3 \pm 5.6$} \\
\hline \end{tabular} \end{center}
$*$ Calculated assuming $\Lsun = 3.844${$\times$}10$^{26}$\,W
\citep{Bahcall++95rvmp} and $\Mbol\sun = 4.75$ \citep{Zombeck90book}.
\end{table}

Using the \Teff s measured in Sect.\,\ref{sec:atm} we have measured the distance to the XY\,Cet system using the method of \citet{Me++05aa}, which is based on calibrations between \Teff\ and surface brightness obtained from interferometry \citep{Kervella+04aa}. An interstellar reddening of $\EBV = 0.04 \pm 0.02$ is needed to obtain agreement between the distances found in the optical ($UBV$) and near-infrared ($JHK$) passbands. The most precise distance measurement is found in the $K$-band, primarily because of its relative insensitivity to reddening and \Teff, and is $279 \pm 6$\,pc. Consistent distances are found using the bolometric-correction technique \citep[e.g.][]{Me++05iauc} and empirical or theoretical bolometric corrections from \citet{Code+76apj} and \citet{Bessell++98aa}. The good agreement between the distances for the two stars in the same passbands is supporting evidence that the light ratio obtained from the light curve is correct.

A comparison of the masses, radii and \Teff s of XY\,Cet to several sets of theoretical models of stellar evolution \citep{Claret95aas,Claret04aa,Girardi+00aas,Pols+97mn} has been performed. We find that the best match corresponds to a solar helium and metal abundance and an age in the region of $850 \pm 100$\,Myr, but the mass--radius relation of the dEB is \reff{less steep} than predicted by any of the models. An enhanced helium or metal abundance can improve the agreement in the mass--radius plane, but at the expense of the agreement in the mass--\Teff\ plane. We conclude that the components of XY\,Cet appear to be discrepant with respect to theoretical predictions, but that the discord is smallest for solar-composition models.


\section{Summary}

XY\,Cet is an eclipsing binary containing two components which show the Am phenomenon. It has been studied many times previously, but without agreement on the radii of the stars and based predominantly on photographic spectra. We have measured the physical properties of the two stars from new phase-resolved CCD spectroscopy and an extensive light curve obtained by the SuperWASP survey for transiting planets. The spectra were analysed using cross-correlation and spectral disentangling techniques, and the light curve was modelled using the {\sc jktebop} code. We measured the effective temperatures of the stars by fitting their H$\gamma$ line profiles, finding good agreement with values from calibration against Str\"omgren indices.

Our spectroscopic data cover only a narrow wavelength region so are not useful for a detailed chemical abundance analysis. Qualitative conclusions have therefore been obtained by comparing the disentangled spectra of the two components to synthetic spectra calculated for the atmospheric parameters of the stars. We find that both stars demonstrate a clear overabundance of iron-group elements and an obvious deficiency of calcium and scandium, as expected for Am stars. Further confirmation of their metallic-lined nature comes from the detection of strongly enhanced abundances of \reff{zirconium and the rare-earth elements lanthanium, cerium and neodymium.}

The physical properties of XY\,Cet are measured to accuracies of 1\% in mass, 2\% in radius and 125\,K in \Teff. We find that the stars are rotating close to synchronicity with the orbital velocity. The system is located at a distance of $279 \pm 6$\,pc. Theoretical models of stellar evolution are able to match the \Teff s of the stars for the measured masses, but predict a steeper slope in the mass--radius diagram than observed.

The physical properties of the component stars of XY\,Cet are not \reff{clearly} different from those of normal A\,stars, supporting the interpretation of the Am phenomenon as a surface effect which does not significantly affect the evolution of stars. The chemical peculiarities and relatively slow rotation of both components mean that XY\,Cet is a promising target for an extensive spectroscopic investigation of the photospheres of metallic-lined stars with accurately known masses, radii and \Teff s.


\section*{Acknowledgments}

\reff{All reduced data from this work will be archived at the {\it Centre de Donn\'ees astronomiques de Strasbourg} and made available at {\tt http://www.astro.keele.ac.uk/$\sim$jkt}.}
JS would like to thank the UK Science and Technology Facilities Council (STFC) for the award of an Advanced Fellowship.
KP acknowledges receipt of the Leverhulme Trust Visiting Professorship which enabled him to perform this work at Keele University.
We thank the referee for a timely and useful report.
The WASP project, funded by the STFC, operates SuperWASP on La Palma and WASP-South at SAAO, and we are grateful to the Instituto de Astrof\'{\i}sica de Canarias and the South African Astronomical Observatory for their ongoing support and assistance
%
%
This paper is based on observations made with the Isaac Newton Telescope operated on the island of La Palma by the Isaac Newton Group in the Spanish Observatorio del Roque de los Muchachos of the Instituto de Astrof\'{\i}sica de Canarias.
The following internet-based resources were used in research for this paper: the NASA Astrophysics Data System; the SIMBAD database operated at CDS, Strasbourg, France; and the ar$\chi$iv scientific paper preprint service operated by Cornell University.


\bibliographystyle{mn_new}

\label{lastpage}

\end{document}